\documentclass[english,aps, preprint]{revtex4}
\usepackage{graphicx}

\makeatletter

\usepackage{babel}
\makeatother
\begin{document}

\preprint{Preprint}

\title{Microwave Assisted Transport in a Single Donor Silicon Quantum Dot}

\author{Enrico Prati}

\affiliation{Laboratorio Nazionale Materiali e Dispositivi per la Microelettronica,
Consiglio Nazionale delle Ricerche - Istituto Nazionale per la Fisica della Materia, Via Olivetti 2, I-20041
Agrate Brianza, Italy}

\author{Rossella Latempa}

\affiliation{Laboratorio Nazionale Materiali e Dispositivi per la Microelettronica,
Consiglio Nazionale delle Ricerche - Istituto Nazionale per la Fisica della Materia, Via Olivetti 2, I-20041
Agrate Brianza, Italy}

\author{Marco Fanciulli}

\affiliation{Laboratorio Nazionale Materiali e Dispositivi per la Microelettronica,
Consiglio Nazionale delle Ricerche - Istituto Nazionale per la Fisica della Materia, Via Olivetti 2, I-20041
Agrate Brianza, Italy}

\email{marco.fanciulli@mdm.infm.it}

\begin{abstract}
Single donors in semiconductor nanostructures represent a key element to develop spin related quantum functionalities in atomic scale devices. Quantum transport through a single Arsenic donor in the channel of a Silicon nano-field effect transistor under microwave irradiation is investigated.
The device is characterized at mK temperatures in the regime of Coulomb-blockade.
Photon assisted tunneling and microwave induced electron pumping regimes are revealed respectively at low and high microwave power. At sufficiently high power, the microwave irradiation induces tunneling through the first excited energy level of the $D_0$ energy of the donor. Such microwave assisted transport at zero bias enhances the resolution in the spectroscopy of the energy levels of the donor.
\end{abstract}

\newpage

\maketitle

\section{Introduction}
\label{sec:1}

Single and double atom impurities in semiconductor nanostructures are of major interest for their key role in atomic scale devices for electronics and spintronics \cite{Kane98,Loss98,DiVincenzo00,Vrijen00,Friesen03,Hanson06}.
The first observation of a single As atom in a Silicon nanostructure \cite{Rogge06,Rogge08} opened the route to the spectroscopy of donor atoms by means of quantum transport measurements. Isolated dopants in the channel of a silicon NanoFET  behave as atomic-like systems with one or two electrons in a central $1/r$ attracting Coulomb potential. Donors can be characterized in terms of energy levels, associated to the $D^0$ and $D^{-}$ states, known in bulk Silicon \cite{Taniguchi76}. A direct measure of the degree of hybridization of the electron wavefunction between the donor potential and the interfacial well has been performed in a Si nanostructure \cite{Rogge08}. The separation of the conduction peaks, observed by applying a gate voltage to a nanostructure, and the threshold voltage, is proportional to the energy level position with respect to the conduction band edge \cite{Rogge06,Rogge08}. 
An isolated, diffused dopant embedded in the Si inversion layer of a nanostructure acts as a quantum dot (Single Donor Quantum Dot, SDQD) at cryogenic temperatures \cite{Rogge06,Sanquer00}. Above the conduction band edge, further quantized energy levels are due to the quantum dot formed in the Si channel by electrostatic confinement.
A microwave field, thanks to the variety of its couplings with the single atom and its environment, provides an unique tool to explore quantum transport and spin properties.  In quantum information processing applications, microwave irradiation should drive electron spin resonance of Zeeman spin doublets and single spin manipulation with pulses of appropriate duration as already observed in GaAs systems \cite{Koppens06}. In Si the observation and the control of single electron spin resonance has not been demonstrated yet due to the small linewidth (expected to be of the order on one gauss) \cite{Silsbee91} if compared to the typical resolution of superconducting cryomagnets ordinarily used for quantum transport measurements, contrarily to GaAs where the spin orbit effect broadens the line to tens of gauss \cite{VonKlitzing83,Prati04,Koppens06}, and to the competition of such effect with photon assisted tunneling in terms of current variation \cite{Obata07}.
When a quantum dot is irradiated with a microwave field, the electromagnetic vector potential provides excitation of electrons in the leads \cite{Kouwen94} and in the dot \cite{Fujisawa97}, and an electromotive potential at the leads \cite{Ferrari05,Prati07,Prati08,Kouwen94,Obata07}. The combination of the two effects goes under the name of photon assisted tunnelling (PAT), previously reported for GaAs split quantum dots\cite{Kouwen94,Oosterkamp98}, and later in a lithographically defined single electron transistor(SET) based on Si/SiGe technology \cite{Dovinos05,Hasko05}. 
Here we show that microwave irradiation can induce and assist transport through a SDQD in Silicon. In Section II, we describe the theoretical framework of microwave effects in nanostructures which applies to the case of the SDQD under investigation. Section III describes the transport characterization of the Coulomb blockade regime of the single As donor. The Sections IV and V present the low and the high microwave power regimes, where PAT and microwave induced electron pumping occur respectively. In the high power regime, excited energy levels participate to microwave induced zero-bias transport. The emphasis is on demonstrating the microwave induced zero-bias spectroscopy as a new tool to improve the donor excited level energy spectroscopy resolution. In Section VI conclusions are drawn on both the power limitations in future electron spin manipulation experiments and the application of the zero bias microwave assisted spectroscopy according to our experimental results.  

\section{Theoretical framework of microwave assisted tunneling for a SDQD}
\label{sec:2}

The effects of the microwave field on a small electron system in a nanostructure have been theoretically studied on the ground of the pioneering work on superconducting junctions of Tien and Gordon \cite{Tien63} by Kouwenhoven in GaAs quantum dots \cite{KouwenPRL94} and later by Williams in Si/SiGe quantum dots \cite{Dovinos05}. The state of the art on photon-assisted transport in semiconductor nanostructures has been summarized by Aguado and Platero in Ref.\cite{Platero04}. 

The NanoFETs under investigation are sketched as three terminal devices where a single donor charge island in the Silicon channel is capacitively coupled to the source (S), drain (D) and gate (G) leads, and highly resistively coupled to the S and D leads only.

Also in a SDQD, according to the theory of electron tunneling in a QD under microwave irradiation \cite{Tien63,KouwenPRL94}, the rf field can be modeled as a potential $\Delta V_{AC}$ cos$\omega t$ added to the static voltage across the leads. Such general ac coupling to three leads can be reduced without loss of generality to only two \cite{Prati07,Prati08}. Following Refs. \cite{Kouwen94} and \cite{Dovinos05} the expression of the tunnel rate through each resistive barrier of the SDQD, used to calculate the current $I_{DS}$, is:
\begin{equation}
\tilde{\Gamma}(\Delta F)=\sum_{-\infty}^{+\infty}J_{n}^2(\alpha)\Gamma(\Delta F + n\hbar\omega) .
\label{RfTunnelRate}
\end{equation}
where $\alpha =\frac{e \Delta V_{AC}}{\hbar\omega}$. In Eq.(1) $J_{n}$'s are Bessel function of the first kind and $\Gamma$ is the tunnel rate in absence of microwave field 
\begin{equation}
\Gamma(\Delta F)= (G/e^2) \Delta F/(1-e^{-\beta \Delta F})
\end{equation}
where $\beta=1/kT$, $G$ is the characteristic conductance of the barrier, and $\Delta F$ is the energy difference between the initial and final states. 

When a microwave field is applied, the additional time-varying potential $\Delta V_{AC}$cos$\omega t$ allows inelastic tunnelling events due to energy exchange between the electrons and photons. Looking at the experimental conditions adopted in the present work, at 300 mK and 40 GHz the quantum regime is achieved since $kT\leq h \nu < \Delta E << U$ where $\Delta E$ is the energy separation of the donor levels which is of the order of few meV (as measured by the Coulomb blockade diamond of the first peak $D^0$), the charging energy $U$ is about 35 meV, and the thermal broadening at 300 mK is $\approx 3 kT = 75 \mu$eV. The intrinsic linewidth of the conduction peaks $\Gamma$ is about 1 meV so $ h \nu < \Gamma$.

It is expected that microwave irradiation induces transport at zero bias ($V_{DS}=0$) in correspondence of the energy levels attributed to the donor, below the conduction band edge. Such microwave induced peaks are expected to exhibit an asymmetric lineshape as a function of the gate voltage $V_G$, due to the random distribution of the electromagnetic field. Such asymmetry reflects the asymmetric coupling of the electromagnetic field with the leads connected to the sample \cite{KouwenPRL94, Dovinos05}. According to the existing model \cite{Dovinos05}, at low microwave power ($\Delta V_{AC} << \Delta E$) the height of the photon assisted tunneling peak has a square root dependence on the microwave power \cite{Platero04}. In the high power regime a linear trend due to the rectification of the ac potentials by the device is also expected to dominate \cite{Ferrari05} so a microwave induced pumping regime is achieved.  Furthermore, for $\Delta V_{AC} \geq \Delta E$, sufficiently high power enables the first excited state to be accessed and a sideband due to the transport through such state is also predicted, as in Ref.\cite{Fujisawa97}

\section{Quantum transport characterization of the single donor}
\label{sec:3}

In order to characterize electronic transport without and with microwave irradiation in a Si NanoFET at 300 mK, we measured the conductance below the conduction band edge, where Coulomb blockade peaks appear. These peaks are due to resonant tunneling through discrete energy levels \cite{Rogge06}. Approximately one sample every ten shows a well isolated pair of peaks which can be related to the energy levels of an As donor, as explained in the following.

The samples were commercial Si NanoFETs with nominal \textit{n}-channel dimensions of 50 nm width and an effective length of less than 70 nm, as revealed by TEM analysis and simulations \cite{Cappelletti}. 
The contacts are doped with arsenic and the channel with boron. Few As atoms can diffuse in the channel. 
From the current-voltage characteristics $I_{DS}(V_G)$ of the device between 8 K and 77 K we measured a threshold gate voltage of 2.850 V. Indeed, below 40 K the isolated conductance peaks become visible but generally it is possible to determine the threshold of the background current down to less than 10 K. The procedure to determine the threshold at low temperature consists to find the threshold at some temperatures and to extrapolate the limiting value.  
The electronic structure of the dot is first characterized by dc transport spectroscopy, i.e. by measuring the Coulomb-blocked oscillations in conductance in absence of any external field. At 300 mK the conductance pattern develops well isolated peaks due to Coulomb blockade (Figure 1). 
In the subthreshold region two current peaks are detected.

We now turn to the identification of such peaks, which, as discussed in the following, are attributed to transport channels opened by a single Arsenic atom levels, as described in Ref.\cite{Rogge06,Rogge08}. We measured the charging energy (energy difference bewteen peaks 1 and 2) $U=E_2-E_1$ and the binding energy $E_C-E_1$, where $E_C$ is the conduction band edge, from the stability diagram shown in Fig. 1.
There, the conductance is plotted in color scale and units of $e^{2} /h$ 
versus gate and bias voltage at 0.3 K. $V_{DS}$ was varied from -7.5 mV to 7.5 mV, while the control gate polarization between 2.2 and 3 V.
Maximum values of conductance are evidenced in white.
The calculated coupling factor, which allows to convert the applied gate voltages to corresponding energies, is $\alpha$=0.14. The charging energy of the donor $U$ is 35 meV, while the spacing between the first peak and the band edge $E_C-E_1$ is 61 meV. Such results are close to those reported by Sellier and coauthors \cite{Rogge06} where $U$=32 meV and $E_C-E_1$=52 meV were observed in a Si FinFET with the source/drain regions implanted with As. The magnetic spectroscopy of the two peaks has been characterized with fields up to 2 T. The peaks linearly shift as a function of the magnetic field $B$ according to the Zeeman splitting $g \mu B$ of a $S=1/2$ spin with $g\approx 2$, as in Ref. \cite{Rogge06}. The first peak shifts towards lower gate voltage values by increasing the magnetic field $B$, while the second peak in the opposite direction. We attribute the two peaks visible below the band edge to the tunneling through an As atom occupied by $N= 0$ (peak number 1) and $N= 1$ (peak number 2) electrons. In other words, the first and second peak correspond to first ($N= 0\rightarrow 1 \rightarrow 0$ ) and second ($N= 1 \rightarrow 2 \rightarrow 1$ ) charge state of the impurity ($D^0$ and $D^{-}$ respectively). 
Details of the first peak ($N= 0\rightarrow 1 \rightarrow 0$ ) are shown in Figure 2 from the the first derivative of the transconductance. The energy scale $\Delta E_{CG}$ refers to an arbitrary zero fixed by the voltage bias of 2.7 V applied to the gate contact. Such gate voltage bias provides a center for the fine voltage sweeps across the first Coulomb blockade peaks. An excited state (ES) energy level marked by a red line is visible at 3 mV from the ground state (GS). The existence of a further excited energy level, marked by a dotted line at about 1.4 meV above the GS, can be observed from the intensity modulation of the transconductance derivative, better shown with a second derivative analysis. The detection of such further excited energy level is significantly improved by the application of the microwave field, as shown in the next sections.

\section{Photon assisted tunneling through the ground state in the low microwave power regime}
\label{sec:4}

The sample was cooled in a 300 mK cryostat while the 1-40 GHz continuous wave radiation was supplied by a 3.5 mm diameter 
berillium in stainless steel coaxial line (UT-141). A dipole antenna was used at the
end of the line to irradiate the sample, located at a distance of about 3 mm. 

The effect of microwave irradiation is clearly visible by measuring the current at zero source-drain bias. 
In Figure 3 a set of experimental $I_{DS}$ ($\Delta E_{cg}$) characteristics of the
first conductance peak is shown for zero bias, where $\Delta E_{cg}$ is the control gate energy shift from the arbitrary center of the sweep, which is $V_{cg}=2.7$ V.
The shape of current oscillations depends on the asymmetry of the coupling between both source and drain reservoirs with the microwave field \cite{Dovinos05}.
Current peaks increase both in amplitude and width as a
function of the power $P_{rf}$, while they do not significantly depend on frequency, as observed in single dots with random microwave field coupling \cite{Dovinos05}. The absolute power at the sample is unknown and changes randomly by several orders of magnitude at different frequencies. Indeed, the electromagnetic environment of the sample strongly depends on the impedance load of the end of the coaxial line which terminates in the $^{3}He$ pot.
In the inset of Fig. 3 a plot of the maximum peak current as a function of the microwave power is shown. The experimental data reveal, at low power, a $c\sqrt{P_{rf}}$ dependence (continuous line) as expected \cite{Dovinos05}.  

\section{Microwave pumping through the SDQD in the high microwave power regime}
\label{sec:5}

We increased the output power of the rf microwave generator and measured the tunnel conductance as a function of its nominal value.
The maximum peak current as a function of microwave power is reported in Fig. 4 for the ($1 \rightarrow 2 \rightarrow 1$) peak.
A double regime of coupling is observed at different power scale. Typical $\sqrt{P_{rf}}$ behaviour \cite{Dovinos05} holds at low power, where photon assisted tunneling occurs, while a linear trend in $P_{rf}$ ( expressed in dBm in Fig. 4) occurs at higher microwave power, where the rectification of the pure voltage fluctuation pumping effect at the leads dominates \cite{Ferrari05,Prati08,Kaestner08,Blumenthal07}. The crossover between the two regimes occurs at about -5 dBm.

We now describe how microwave irradiation may provide an accurate tool for the spectroscopy of excited states at zero bias. Analogously to the spectroscopy of isolated atoms, microwaves can be used as a probe to electrically reveal internal excitation energies in atomic-like systems, such as SDQDs.

In Fig. 5 a colour plot of the transconductance $\partial I_{DS}/\partial (\alpha eV_{cg})$ variation in the plane (${P_{rf}}$, $E_{cg}$) is reported for the ($0 \rightarrow 1 \rightarrow 0$) transport.  Derivative maxima are evidenced in white, while dark zones refer to minima. Conductance variations develop a triangular sector evidenced by black lines with slopes following the positive (left, white zone) and negative (blue, right zone) values. 
The bold vertical lines mark the energy values where the maximum of conduction occurs. In correspondence of a nominal power of about -2 dBm, an excited state is detected at an energy 1.4 meV higher than the ground state GS at energy $E_1$. 

The second state shown in Fig. 5 is attributed to an excited state of the donor, accessed at zero bias only in the high power regime, when the voltage oscillations of the leads increase in amplitude. 
Figure 6 shows the modified stability diagram of the $D^{0}$ peak in the proximity of zero bias when the sample is irradiated. The frequency of the microwave source was set at 40 GHz and the power at 0 dBm. At such microwave power, both the GS and ES were resolved separately. The inversion of the sequential tunneling current at the zero bias line (obtained by setting the bias voltage $V_{DS}=-0.5$ mV because of an offset) is dominated by the microwave assisted current in the $V_{DS}$ region between -0.5 mV and -1.5 mV. Therefore the total current remains positive and it shows a double peak. At $V_{DS}$ voltages below -1.5 mV, the current inverts its sign and exhibits a double negative peak in correspondence of the two energy levels.

The existence of such excited state at 1.4 meV state is confirmed by the ordinary stability diagram (Figure 2).  However, due to the small energy scale associated to the microwave frequency (of the order of $h\nu$), zero bias microwave  spectroscopy (ZBMS) takes advantages of a much higher resolution when compared with the method of stability Coulomb blockade diagrams to isolate excited states. Indeed, only a second derivative study of the stability diagram allows the clear discrimination of such excited level at 1.4 meV, because of the large width of the Lorentzian peak $\Gamma$ of the order of 1 meV. The conversion factor $\alpha$ used for the microwave induced gate fluctuations is the same used for dc voltages since it only depends on the capacitances of the island with respect to the gate, source and drain contacts, which are independent from the voltage source, at least in the range of the frequencies used in our experiments. 
The superior resolution reached by means of the ZBMS is shown in the Figure 7. The current at $V_{DS}=2.8$ mV with and without the microwave irradiation at 40 GHz and 0 dBm is compared with the current at almost zero bias ($V_{DS}=0.3$ mV) due to the microwave effect, for which the two peaks are clearly separated. 

The high resolution is originated by the mechanism represented in Figure 8: the microwave induced voltage oscillations in the contacts may be reduced from three to two, for example the drain and the gate contacts, and their relative phase can only be $\phi=0,\pi$.\cite{Ferrari05} The relative coupling ranges between the limiting case 1 and 2 of the Figure 8 (left side). While the case 1 is not capable to account the high resolution, the case 2 provides the condition to realize the preferential tunneling through the ES in opportune condition. 
In case 2 the microwave field provides an oscillation of energy amplitude $e \Delta V_{GS} (P_{rf})$  and it holds $e \Delta V_{DS} (P_{rf}) << e \Delta V_{GS} (P_{rf})$. 
According to the threshold behaviour observed in the experiment, in the case (2) two different behaviour are predicted when the working point of the device is in the region $N=1$ below the ES energy, depending on the amplitude of the oscillations of the quantum dot energy levels connected to the gate. When the energy oscillation in the quantum dot $e \Delta V_{GS} (P_{rf})$ is lower than the energy gap between the GS end ES ($\Delta E_{01} = E_1 -E_0$), only GS participate to transport. On the contrary, when $\Delta E_{01} \approx e \Delta V_{GS} (P_{rf})$, the eletrons are pumped from the left to the right, as shown in the right side of the Figure 8. In this case the GS is initially (a) filled with an electron in the region $N=1$. The drain coincides with the left contact. After the first quart of period (b) the working point reaches the region $N=0$ of the stability diagram and the electron is pumped on the left side by virtue of the small unbalance between the Fermi energy in the left and the right contacts, with the eventual participation of photons which assist the transport. In the next half period the working point moves back to the region $N=1$ (c) and may reach the point (d) with the quantum dot still empty. It is crucial to observe that the tunneling from the left contact to the ES is more probable than that to the GS, since for the first $d \Delta V_{GS}/ dt \approx0$ while for the second $d \Delta V_{GS}/ dt$ is maximum, so the transition has more time to occur through the ES.
The model predicts therefore that for $e \Delta V_{GS} (P_{rf}) >> \Delta E_{01}$ the high resolution is lost, which is confirmed by the experiment (see Figure 5 at high power). The period ends in position (e) with the quantum dot empty. The dot is therefore filled from one of the two contacts with the same probability, so the net current on several periods is from the left to the right. 

\section{Conclusion}
\label{sec:6}

Microwave irradiation induces transport in Single Donor Quantum Dots with a complete analogy to split gate confined and lithographically defined quantum dots. Two different regimes of coupling have been observed, the former caused by photon assisted tunneling, and the latter by quantized charge pumping effects respectively. Microwave irradiation, at sufficiently high power, allows the detection of excited states at zero bias with a resolution which is much higher than the one obtained by means of standard stability diagrams. 

We explained the high resolution of the zero bias microwave spectroscopy by means of a model capable to explain its three fundamental characteristics, namely its dependence on the distribution of the microwave field on the contacts, the threshold behaviour above a certain microwave power, and the disappearing of the effect at high power. All the effects reported in this work impose a strong limitation on the microwave intensity to be used for future donor-based quantum device applications, where single electron spin dependent currents should be discriminated. The maximum microwave power should be limited by the spin effect current fluctuation expected signal, which imposes a tradeoff between the minimum power required to produce spin reversal, and the maximum power limited by the appearence of photon assisted tunneling. On the other hand, the zero-bias microwave spectroscopy can been exploited to reveal spectroscopic features not otherwise observable with the same resolution using more conventional techniques.

\newpage

\begin{acknowledgments}
The authors thank Dr. Cappelletti (STMicroelectronics) for providing the samples. 
\end{acknowledgments}
 
\section{Figure Captions}
Figure 1 Stability diagram in the subthreshold region. Two clearly isolated peaks lies below the conduction band edge. The charging energy $U$ is 35 meV, while the difference $E_C-E_1$ is 61 meV, in agreement with As donors in Si nanostructures as  reported in Ref. \cite{Rogge06,Rogge08}. The conductance has been calculated by dividing the measured current $I_{DS}$ by $V_{DS}$, so the data at $V_{DS}=0$ are not defined.

Figure 2 Differential transconductance of the first peak $N= 0\rightarrow 1 \rightarrow 0$ with $eV_{DS}$ between -10 meV and 10 meV. A well resolved excited state (ES) (red line) at 3 meV above the ground state (GS) (black line) is detected. A further excited level at 1.4 meV (red dotted line) above the GS can be distinguished as a weak variation of the first derivative. Its resolution is highly magnified by the microwave irradiation as discussed in the following.

Figure 3 Current-Voltage characteristics of the first conductance peak at zero bias and base temperature of 300 mK for different rf powers. The control gate voltage $V_{cg}$ is converted into the corresponding energy shift $\Delta E_{cg}$ from the arbitrary center of the sweep, positioned at 2.7 V. Inset: the square root dependence of the peak height, typical of photon assisted tunneling \cite{Dovinos05}. The bias voltage $V_{DS}$ was set to compensate the instrumental small bias offset, so that the current peak was comparable with the noise level, and the observed current is activated only by the microwave irradiation.

Figure 4 Maximum current peak as a function of nominal applied microwave power on the second peak. Depending on the applied power, the regime changes from a square root trend $P_{rf}^{1/2}$ (low power) where photon assisted tunneling dominates, to a linear trend $P_{rf}$ (high power), where the voltage fluctuation in the leads is the only responsible of electron pumping.

Figure 5 Transconductance variations as a function of microwave power of the first peak ($D^{0}$ state). Above -2 dBm the power is sufficient to access an excited state (ES) 1.4 meV higher than the ground state (GS) energy $E_1$.

Figure 6 The stability diagram of the first peak ($D^{0}$ state) when the sample is irradiated with a 40 GHz microwave. The microwave power is set at 0 dBm where the ac voltage oscillation in the leads was observed to allow the tunneling through either the GS and the ES as a function of time along a period.
 
Figure 7 The microwave irradiation may lead to observe the fine details of the spectrum in the vicinity of zero bias condition. Here the excited state at 1.4 meV is observed by applying a bias energy $eV_{DS}=0.3$ meV  at a microwave frequency of 40 GHz with nominal power of 0 dBm. Such microwave assisted current at zero bias is compared to the current observed at $eV_{DS}=2.8$ meV with and without the microwave irradiation. 

Figure 8 Scheme of the mechanisms enhancing the spectroscopic resolution by means of microwave irradiation in zero bias condition.  On the left side the energy orbits in two limiting cases are drawn on the stability diagram of the $(0\rightarrow 1\rightarrow 0)$ peak at $V_{DS}=0$. In the example (1) $e \Delta V_{DS} (P_{rf}) >> e \Delta V_{GS} (P_{rf})$ (with phase $\phi=0$) and no enhancement is possible by sweeping the dc gate voltage. In the example (2) $e \Delta V_{GS} (P_{rf}) >> e \Delta V_{DS} (P_{rf})$ (the opposite slope sign indicates that $\phi=\pi$). In the latter case, the isolated secondary peak corresponding to the $ES$ appears when the gate voltage is approximately set in correspondance of the ES peak by the following steps (right side): a) the quantum dot is originally in the $GS$ with no bias applied; b) the small microwave induced imbalance of the drain causes the shift of the electron to the left side, eventually assisted by photons; c) the GS and the ES are pushed to energies lower than the Fermi energy $E_{Fleft}$ of the left lead. The transition probability to the ES is higher than the probability to the GS. Indeed, the sine function has maximum time derivative when $E_{0}\approx E_{Fleft}$ while it vanishes when $E_{1}\approx E_{Fleft}$; d) the opposite imbalance of the point b) moves the charge to the right side. e) when the period has been completed, the dot is left empty and the ground state is filled again with approximately equal probability from one of the two leads.

\newpage

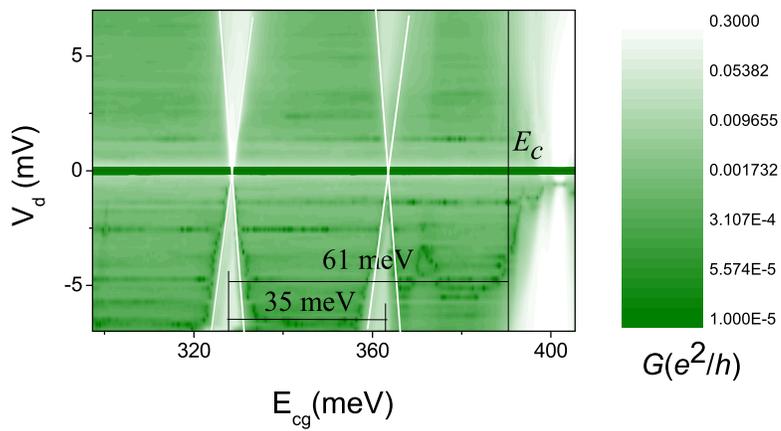
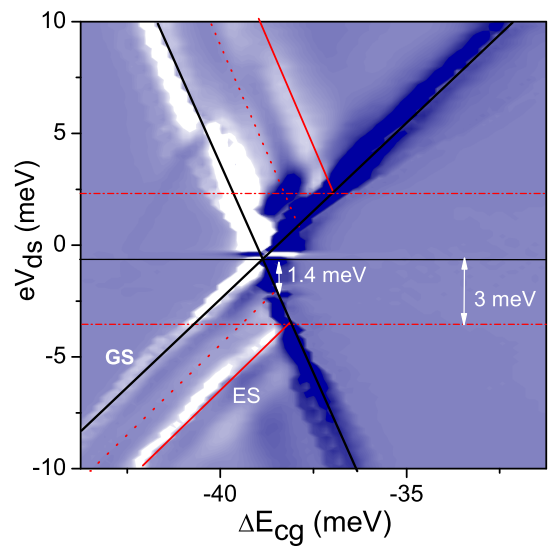
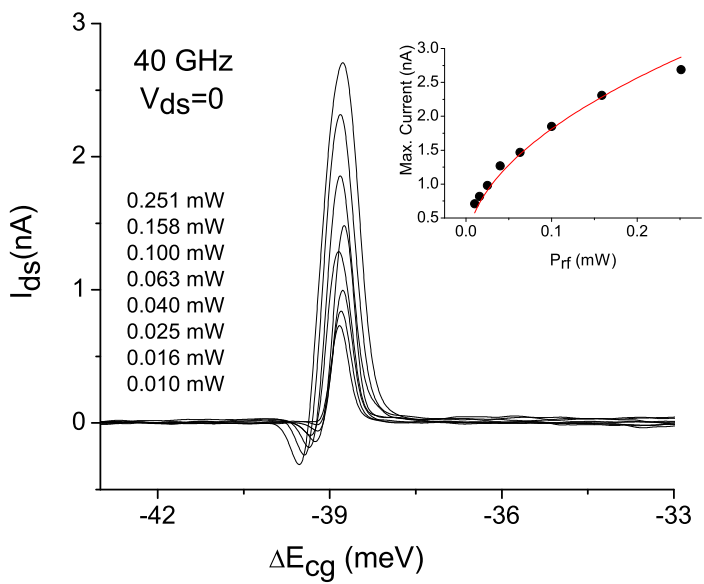
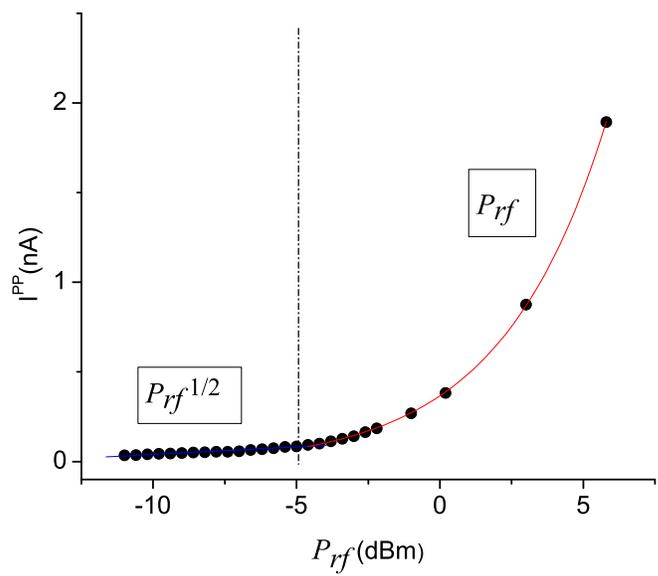
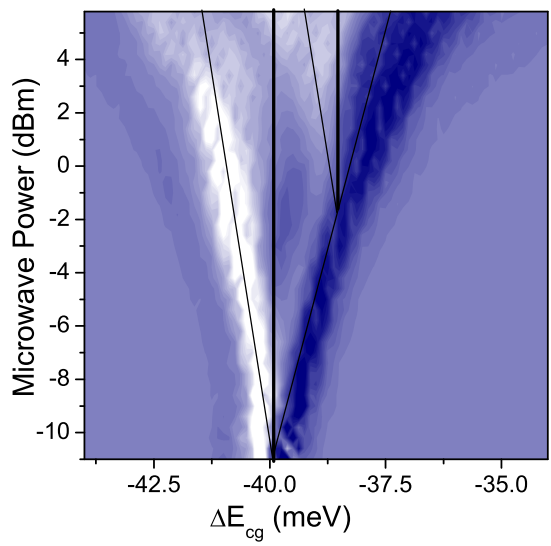
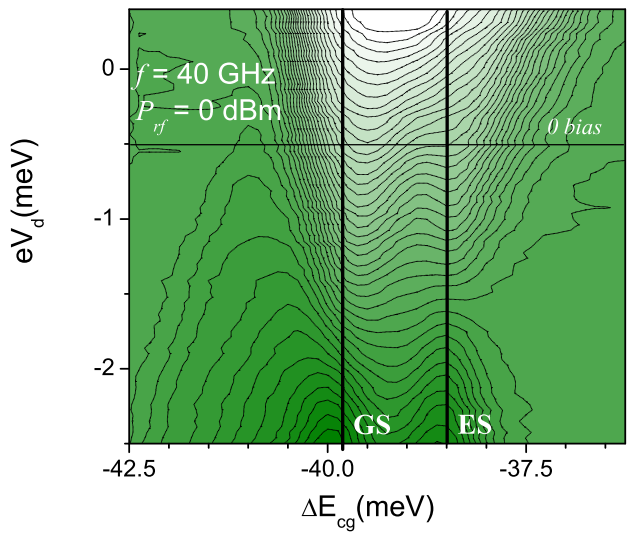
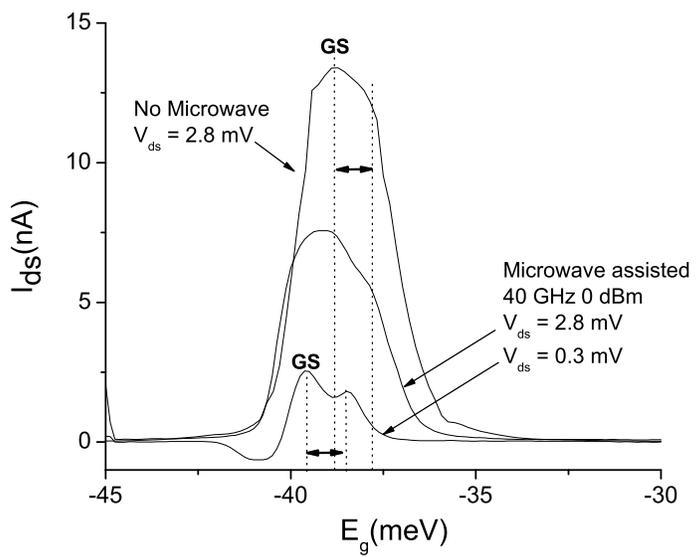
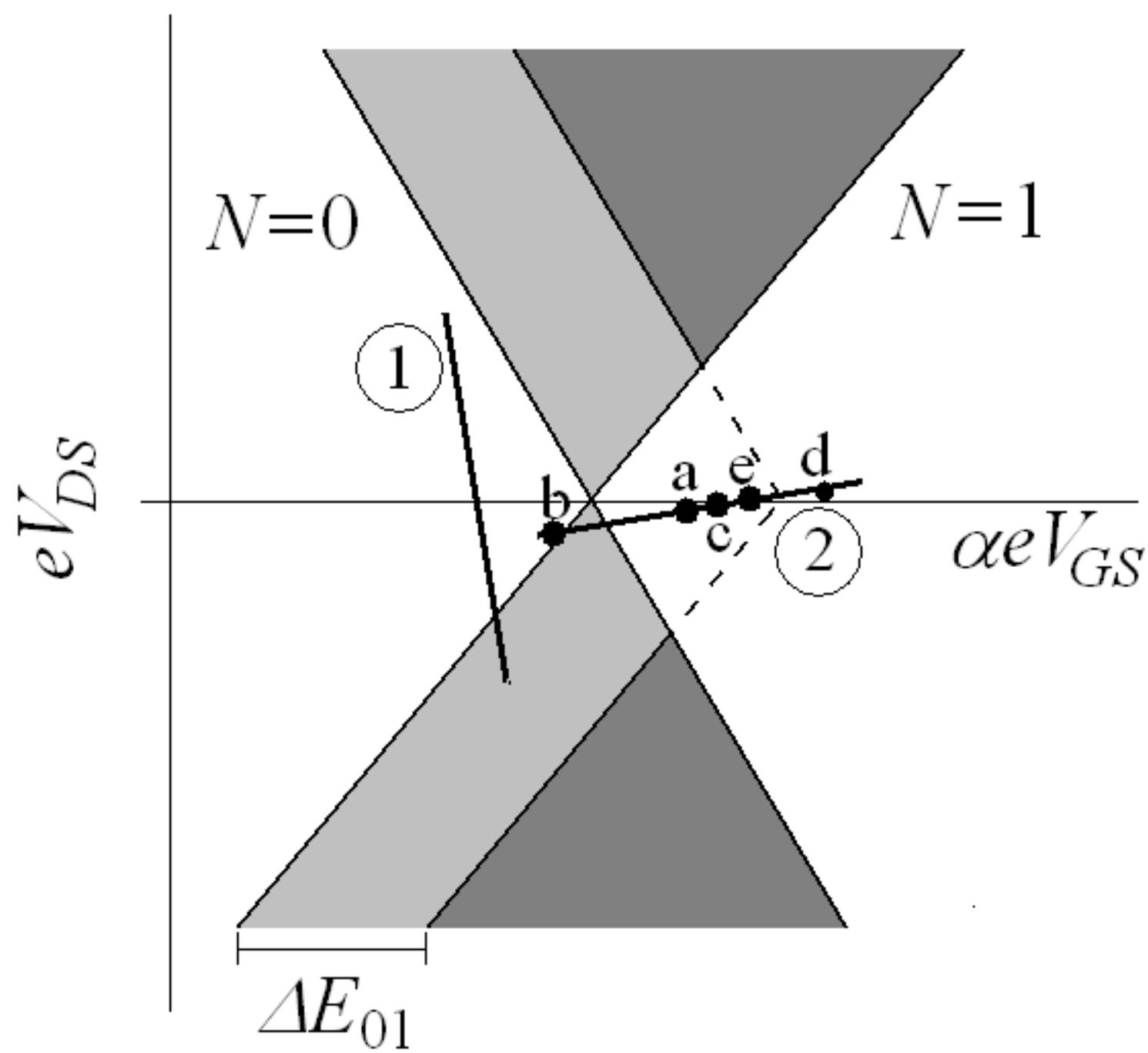
\end{document}